\documentstyle[epsfig]{texas}

\newcommand \beqas{\begin{eqnarray*}}
\newcommand \eeqas{\end{eqnarray*}}

\begin{document}

\title{NEUTRON STAR STRUCTURE}

\author{C. J. PETHICK$^{1,2}$, A. AKMAL$^{3}$, V. R. PANDHARIPANDE$^{2}$, and\\
D. G. RAVENHALL$^{2}$}

\address{(1) NORDITA,\\
Blegdamsvej 17, DK-2100 Copenhagen {\O}, Denmark}

\address{(2) Department of Physics, University of Illinois at
Urbana-Champaign,\\
1110, West Green Street, Urbana, Illinois 61801-3080, USA}

\address{(3) NASA/Goddard Space Flight Center, \\Code 682.3, 
Greenbelt, MD 20771, USA}

{\rm Email: pethick@nordita.dk, aakmal@cfa.harvard.edu, vijay@uiuc.edu,\\
\hspace*{1.6cm} ravenhal@uiuc.edu}

\begin{abstract}

A review of properties of matter in the interior of neutron stars is given.  
Particular attention is paid to recent many-body theory calculations of the 
properties of dense matter. Among topics discussed are the strong increase of 
tensor correlations at relatively low densities, the ``relativistic boost term" 
in the interaction, and the sensitivity of properties of neutron star models to 
three-body forces. 
\end{abstract}

\section{Introduction}

Before turning to theory, it is salutary to consider what information about
neutron stars may be learned from observations.  One basic quantity is the
stellar mass, and an important recent development is the tightening of the
mass estimates for neutron stars in binary systems by Thorsett and Chakrabarty
for radio pulsars \cite{tc}, and for Chakrabarty and
Thorsett for accreting neutron stars \cite{ct}.  In addition, as we shall
hear from van der Klis later in this symposium \cite{vdk}, (see also 
Ref. \cite{lmp}), an exciting possibility of obtaining
limits on masses of neutron stars from measurements of the frequencies of
kilohertz quasiperiodic oscillations has recently arisen.  Thus information
about masses of neutron stars in binaries is becoming increasingly precise,
and provides important constraints on models of neutron stars.

While a knowledge of the equation of state is the only input required for
determining the mass-radius relationship for neutron stars, and hence the
range of physically allowable neutron star masses, there are a number of
observational phenomena of neutron stars that depend on other
properties of matter in stars, apart from the equation of state.
Among these are the surface emission from neutron stars, which will be
discussed later in this symposium by Caraveo \cite{caraveo} and Shearer 
\cite{shearer}, and glitches.

In this talk we shall discuss a number of aspects of neutron star
behaviour, and will describe some recent calculations of the equation of
state of neutron star matter and the neutron star models that these
equations of state predict.  To set the scene, we give in Fig.~ 1
\begin{figure}
\centering
\epsfig{figure=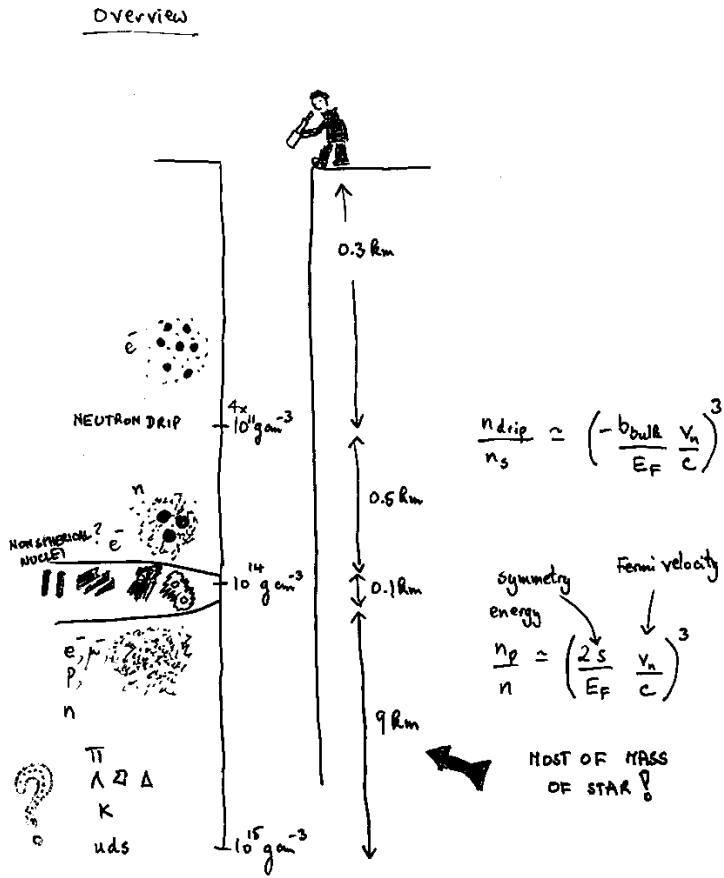, width=10cm}
\caption{An overview of the structure of a typical neutron star.
On the left are sketches of the kind of matter expected at the 
indicated densities, and on the right are simple approximate
expressions for relationships between these densities.
}\label{fig1}
\end{figure}
an overview of the structure of a neutron star.  At the lowest densities
matter is similar to terrestrial matter.  Electrons become increasingly
free as the density increases, and at a density of about 10$^6$ g/cm$^3$
they become relativistic.  The increasing electron Fermi energy makes it
energetically favorable to reduce the number of electrons, and convert a
corresponding number of protons into neutrons.  Consequently with
increasing depth in the star matter becomes more neutron rich, and at a
density of about $4 \times 10^{11}$ g/cm$^3$ matter becomes so neutron rich that
the most energetic neutron states in nuclei become unbound, and begin to
occupy continuum states.  This is neutron drip.  As the density increases
further, the density of neutrons between nuclei also increases, and
eventually when the density of neutrons outside nuclei is roughly the same
as that of the matter in nuclei, nuclei merge to form a uniform fluid
consisting mainly of neutrons, with a few percent of protons, together with
electrons to ensure the electrical neutrality of the matter.  At densities
just below that at which nuclei merge, one expects nuclei to be aspherical,
having the form of extended rods and plates \cite{leshouches}.

Above nuclear density the physics of matter is more uncertain, but
the equation of state and composition of matter at these densities are
central ingredients in constructing neutron star models and predicting
their evolution.

At the lowest densities, properties of matter may be predicted directly on
the basis of measured nuclear data, since up to just
below neutron drip density the nuclei predicted to occur are ones
accessible to laboratory study.  For the higher density region up to just
above nuclear density, the basic constituents of matter are neutrons and
protons, together with the electrons.  The interaction of two
nucleons is rather well understood at the energies of relevance, and the
system is sufficiently dilute that the methods of many-body theory
may be used with confidence to predict the properties of matter.
Above nuclear density the situation is less clear.  Basic questions are
what the fundamental constituents are, what their interactions are, and how 
the energy of dense matter can be calculated.  These questions are not
independent, since whether or not a particular constituent is present
depends on what its energy is.

There are basically two sorts of approach to calculating the properties of
dense matter.  The first is to take an interaction between constituents
which is as realistic as possible, and fits known scattering data, and to
then use the techniques of many-body theory to calculate correlations in
the matter, from which one can then evaluate the energy and the equation of
state.  The second is to adopt a schematic model, often of the mean field
type, where the coupling stengths for the various interactions are treated
as parameters to be fitted to a limited set of observable quantities.  Both
approaches have their advantages and disadvantages.  As far as the
many-body approach is concerned, the nucleon-nucleon interaction is well
understood, and the methods of many-body theory are well developed.
However, the approach is generally based on non-relativistic theory.  Some of
the schematic models are better able to incorporate the effects of relativity,
and can be easily generalized to allow for the presence of many
constituents.  The disadvantage of the approach is that the complicated
correlations in dense matter, which are very density dependent, are
represented by a number of parameters giving the strengths of the various
mean fields.  Both approaches, however, are hampered by the paucity of
relevant experimental data.  While pair-wise interactions of nucleons are
well understood at low energy, three- and higher-body interactions of
nucleons are not well characterized, and even information about two-body
interactions involving constituents other than nucleons is very limited.

\section{Basic considerations}

In this talk we shall describe some recent calculations of
the properties of dense matter from many-body theory \cite{apr}.
The wave
function, $\Psi$, for the many-nucleon system, whether it be pure neutron
matter
or isospin-symmetric nuclear matter with equal numbers of neutrons and
protons, is assumed to be of the Jastrow form,
\begin{equation}
\Psi=\prod_{i<j}f_{ij}\Psi_0,
\label{jastrow}
\end{equation}
where the indices $i$ and $j$ label particles, $\Psi_0$ is a Slater
determinant for a gas of free nucleons, and $f_{ij}$ is a correlation
factor, which depends on the distance between two particles, and other
variables such as the  spin and isospin of the interacting pair of
nucleons.  The procedure is to evaluate the energy of the system with the
wave function Eq. (\ref{jastrow}).

Information about two-body nucleon nucleon interactions is derived from the
Nijmegen data set for nucleon-nucleon scattering at energies below 300 MeV
in the center of mass frame.  This was fitted by Wiringa, Stoks
and Schiavilla \cite{wss} to a non-relativistic
interaction, denoted by Argonne v18 (or simply A18), which allows for isospin 
dependent
effects, such as those coming from the mass difference between neutral and
charged pions.  The ``18'' denotes the number of operators in the effective
two-body interaction.

It has long been known that the two-body interaction alone is insufficient
to account for the binding of nuclear matter and of light nuclei.  This
discrepancy has been attributed to the presence of three-body forces, in
which three nucleons interact simultaneously.  This takes into 
account non-nucleonic
degrees of freedom, which are not present in the wave function
(\ref{jastrow}).  One contribution to this interaction arises from two
nucleons scattering, with one of the nucleons being excited to a $\Delta$
resonance state, which subsequently de-excites by scattering from a third
nucleon.  Experimental information about the three-body force is sparse,
and in the calculations it is represented by a simple expression which has
the theoretically predicted behaviour at large distances but with a
strength to be obtained by fitting to experiment, and a parametrized form
for the short-range part.  Parameters in the interaction are adjusted to
account for the binding energy and density of light nuclei and nuclear
matter.

A novel feature of the present set of calculations is the inclusion of the 
so-called
``relativistic boost correction''.  As we mentioned above, the methods of 
many-body theory are based on 
non-relativistic concepts, but it is possible to include the leading
effects of relativity by appropriate
modification of the effective interaction.  Such a formalism is familiar in
the context of electrodynamics, where it leads to the so-called Breit
interaction between electrons.  Within the context of the nuclear many-body
problem its importance has been appreciated for more than a quarter of a
century \cite{boost1}, but it has generally not been included in previous 
calculations \cite{boost2}.

The effective Hamiltonian thus has the form
\begin{equation}
H_{NR} = \sum_i \frac{p_i^2}{2m} + \sum_{i<j}\left(v_{ij} +\delta
v^b_{ij}\right)  +  \sum_{i<j<k}v_{ijk},
\label{ham}
\end{equation}
where $v_{ijk}$ is the effective three-body interaction, and $
\delta v^b_{ij}$ is the relativistic boost correction.  The latter has the
form
\begin{equation}
\delta v^b({\bf P})=-\frac{P^2}{8m^2}v + \frac{1}{8m^2}[{\bf P.r} \; {\bf
P.\nabla},v]+ ({\rm spin-dependent\; terms}).
\label{boost}
\end{equation}
In this expression $\bf{P}$ is the relative momentum operator for the pair of
particles and the $\nabla$ operates on the relative coordinate ${\bf r}$ 
for the two
particles, and the two-body interaction is written simply as $v$.

The most complex part of the calculations is the evaluation of the energy
for the trial wave function (\ref{jastrow}).  We shall not describe
the technicalities here, but will go directly to a discussion of some of
the salient features of the results.

First of all we consider results for light nuclei  $^3$H and
the alpha particle, $^4$He.  The boost correction to their energies
amounts to almost 40\% of the difference between the experimental energies
and the theoretical ones calculated using the two-body interactions alone.
Consequently, if the boost correction is included, the strength of the
three-body interaction required to produce agreement between theory and
experiment is about 40\% less than that required if the boost correction is
neglected.

\section{Strong tensor correlations}

For uniform matter, a striking discovery is a strong increase in tensor
correlations at relatively low densities.  For pure neutron matter
this occurs at a density of $\sim 0.2$ fm$^{-3}$, just above nuclear matter
density ($\sim 0.16$ fm$^{-3}$), while for symmetric nuclear matter it occurs at
a density of $\sim 0.32$ fm$^{-3}$.  For pure neutron matter these tensor
correlations reflect a tendency for the spin components
perpendicular to the vector joining two neutrons to be aligned, while
the components along the vector joining the spins tend to be anti-aligned.
Such correlations arise from the tensor force, whose dependence on angle is
given by $3 \sigma_1 \cdot \hat {\bf r}_{12} \; \sigma_2 \cdot \hat {\bf r}_{12}
-\sigma_1 \cdot \sigma_2$.  If such correlations were of long range,
neutron matter would have a neutron spin density wave, or in other words
it would be a layered antiferromagnet.  Such a state, if described in terms of
pionic degrees of freedom, would be a Bose condensation of neutral pions.

This result is noteworthy because for the past two decades, the accepted opinion
has been that pion condensation was ruled out by strong repulsive central
correlations which suppressed the tensor ones.  However, in view of the
recent results the possibility of pion condensation should be re-examined.
The presence of the stronger pionic correlations results in the energy of
dense matter rising less rapidly with increasing density than one would
anticipate on the basis of calculations in which the tensor correlations
are assumed to be insensitive to density.

Next we consider how predictions for the energy of dense matter are
influenced by the inclusion of the boost correction.  This is exhibited in
Fig.(2).  
\begin{figure}

\centering

\epsfig{figure=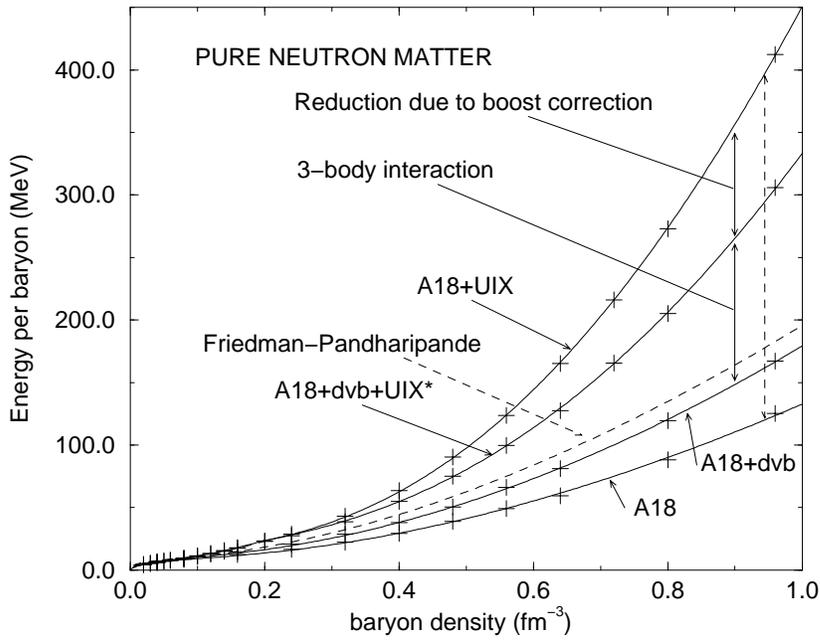, width=12cm}

\caption{Energy per baryon of pure nuclear matter 
as a function of density for various
nuclear models: A18 is the Argonne v18 two-body interaction; $\delta v^b$ is the
relativistic boost correction to the energy; UIX refers to the
3-body interaction with parameters determined by fits to light nuclei and 
nuclear matter in the absence of 
the boost correction, while UIX$^*$ is the corresponding interaction 
fitted with the boost correction 
included.
}\label{fig2}

\end{figure}
If the boost correction is included, with a three-body
interaction adjusted to give a fit to the properties of light nuclei and
nuclear matter, the energy rises less rapidly than if one neglects the
boost correction and uses a correspondingly stronger three-body interaction
so that the properties of observed nuclei are reproduced.  The reason for
this effect is that the contribution to the energy of matter coming from
the boost correction increases less rapidly with density than does that
coming from three-body interactions.   At a density of  $\sim
0.8$ fm$^{-3}$ the contribution from three-body interactions is reduced by
about a factor of two if the boost correction is included.  In view of the
uncertainty of the form of the three-body interaction, this reduced
sensitivity of results to the three-body interaction is encouraging.

It is interesting to compare the results of calculations based on
variational methods with ones based on Brueckner theory \cite{engvik}.  
Within the
Brueckner approach it is difficult to evaluate higher order correlations,
and most calculations are performed at the two-body level.  Such
comparisons have been made neglecting the boost correction and three-body
terms in the interaction, and for neutron matter one finds remarkably good
agreement between the two approaches.  This is reassuring, because it indicates
that the effects of high-order correlations is not overwhelming, probably
due in part to the fact that for pure neutron matter, the Pauli principle
prohibits three particles from being at the same point in space.  For
symmetric nuclear matter short-range correlations are not suppressed by the
Pauli principle, and the variational and Brueckner results differ
increasingly for densities above a few times nuclear density \cite{refapr}.
A further point to notice that there are a number of different
two-nucleon interactions based on the Nijmegen data set, but where comparisons
have been made, results do not depend on which particular form of the
interaction one uses, so long as it fits the Nijmegen data.

\section{Other constituents}

The next topic we address is the possibility of other constituents being
present in matter.  We have already mentioned pion condensates, and among
the others are a kaon condensate, hyperons such as $\Sigma^-$ and
$\Lambda$, and quark matter, in which hadrons are crushed and the basic
degrees of freedom of the system become more or less free quarks.  The
appearance of new constituents in neutron stars is of importance for two
reasons.  First, if it becomes energetically favourable to exploit a new
degree of freedom, the energy will be reduced compared with the energy of
the system in which that degree of freedom is suppressed.  In other words,
the energy will rise less rapidly with density, a fact which implies a
general softening of the equation of state, and a reduction in the maximum
mass of a neutron star.  Second, essentially all possible new degrees of
freedom of the system that have been considered to date allow for neutrino
emission to occur via the direct Urca process.  This is much faster than
the modified Urca process, which is the standard process for matter
consisting of neutrons and a small fraction of electrons and protons.  This
is expected to have far-reaching consequences for the thermal evolution of
neutron stars. The basic condition for a constituent of charge $Ze$ and
baryon number $A$ to be present in equilibrium is that its chemical
potential is given by
\begin{equation}
\mu(A, Z)= -Z\mu_e +A\mu_n,
\label{mu}
\end{equation}
where $\mu_e$ is the electron chemical potential and $\mu_n$ the neutron
one.  In Fig.(3) 
\begin{figure}

\centering

\epsfig{figure=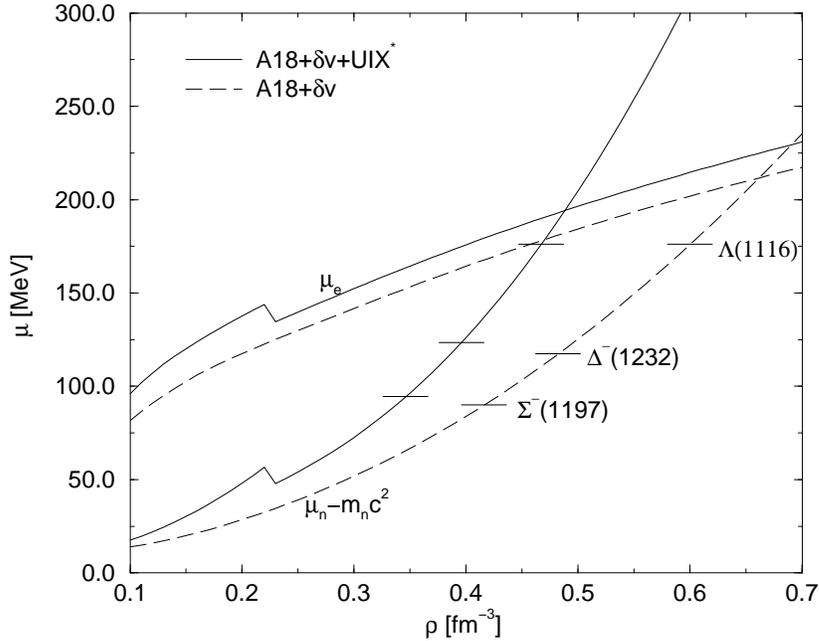, width=12cm}

\caption{Neutron and electron chemical potentials in beta stable 
matter for the two models $A18+dvb$ and $A18+dvb+UIX^*$. 
Threshold densities for the appearance of non-interacting
hyperons are marked by horizontal line segments.
}\label{fig3}

\end{figure}
we show plots of $\mu_n$ and $\mu_e$ as functions of
density for matter in beta equilibrium.  The bars denote the lowest density
at which particular constituents appear {\it if they are treated as
non-interacting}.  Observe how the inclusion of the three-body interaction
increases the neutron chemical potential, and thereby lowers
the threshold densities for other baryonic constituents.  The fact that
interactions of the possible new constituents have been neglected here
means that the threshold estimates given are unrealistic.  One further
observation is that the model with the boost correction and three-body
interactions included predicts proton concentrations high enough for the
direct Urca process for nucleons to occur at densities in excess of $\sim
0.75$fm$^{-3}$.  However, such densities are expected to occur only in the
cores of stars with masses greater than about 2$M_{\odot}$.

\section{Equation of state and stellar models}

The pressure may be calculated from the energy density $E(\rho)$, where
$\rho$ is the number density of baryons, and the properties of neutron
stars may then be obtained from the conditions for hydrostatic equilibrium in
general relativity, the Oppenheimer-Volkoff equations.  The mass-radius
relationship that results is shown in Fig.(4), indicating a maximum mass of
about 2.2M$_{\odot}$.

\begin{figure}

\centering

\epsfig{figure=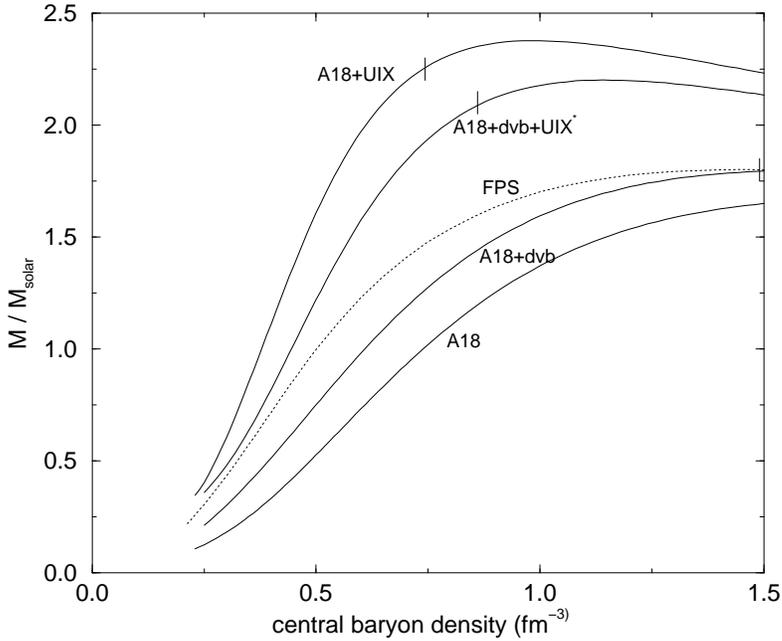, width=12cm}

\caption{Neutron star mass versus radius for the nuclear
models discussed in this paper compared with the results of the older FPS
interaction based on the results of Ref. (10).  UIX denotes the three-body 
interaction fitted to nuclear 
properties if the relativistic boost correction is neglected, and UIX$^*$ is the 
corresponding interaction 
if the boost correction is included.}
\label{fig4}

\end{figure}

One unfortunate feature of the equation of state obtained in these
calculations is that matter is superluminal, that
is the sound speed, given by $c_s = (\partial P/\partial \tilde
\rho)^{1/2}$ exceeds the speed of light.  However, this occurs only at
densities greater than about 1 fm$^{-3}$, which would be realized only in
the centres of stars close to the upper mass limit.  This pathology
reflects the fact that the theory is not completely consistent with special
relativity.  Two sorts of approach, short of developing a complete theory
consistent with special relativity, can be adopted to remedy this defect.
One is to assume that the equation of state is known reliably up to some
density, and then assume that at higher densities the sound velocity is
equal to $c$, the maximum value consistent with relativity.  Such a
procedure, first used by Rhoades and Ruffini \cite{rr} and more recently by
Kalogera and Baym \cite{bk}, leads to an equation of state which is stiffer
than is realistic, and thus gives larger stellar masses for a given
central density.  Another approach is to argue that physically it is a
better approximation to use at high densities a model of matter that is based 
on degrees of freedom other than the nucleon ones that are appropriate at low
densities.  The particular example we consider is quark matter.  The
question to be addressed is then how to interpolate between the high and
low density limits.  This is a complicated problem since in the region
where the crossover from one state to another occurs neither the low- nor
high-density equations of state are expected to be good approximations.
Putting this problem aside, the simplest way to make a thermodynamically
consistent description of the crossover is to adopt the standard method of
making a Maxwell construction, which leads generally to a first-order phase
transition at which the density of matter jumps discontinuously in the
star.  Glendenning pointed out that another solution might be possible
because there could be an intermediate phase consisting of coexisting quark 
and nucleon phases in a uniform background of electrons \cite{glen}.  The
regions of quark matter and nucleons should have length scales small
compared with the electron screening length, and consequently the quark and
nuclear matter regions need not be electrically neutral locally, even though 
they
would be neutral on length scales large compared with the electron
screening length.  The lack of local charge neutrality implies that the
pressure need not be constant in the intermediate region, as it is if
matter is locally electrically neutral.  Which of the two ways of matching
the two equations of state is closer to the truth depends on the surface
tension between the two phases:  if the surface tension is too high the
length scale of the quark and nucleon regions will exceed the electron
screening length and the assumption of a uniform background of electrons
will fail.  In this case the Maxwell construction for two phases of neutral
matter would give the correct answer, while for a lower surface tension,
the mixed phase would be a possibility.  Assuming the mixed phase and
neglecting surface energies and the Coulomb energies associated with the
charge inhomogeneities clearly underestimates the energy of the system, and
thus this procedure is expected to lead to the ``softest" equation of state
in the intermediate region.  However, it must be borne in mind that these
methods of describing matter at intermediate densities are just devices
behind which we try to hide our basic ignorance about what is really going
on there, and any phase transitions that are found in model calculations
may be purely due to the inadequacy of our fundamental understanding.

Let us now consider the consequences of interpolating between the nucleon
equation of state described earlier and one for a free gas of quarks at
high densities, assuming that the equation of state in the intermediate
region is calculated on the assumption of the coexistence of regions of
nucleon and quark matter with different charge densities.  First of all,
matter is never superluminal for the values of the bag constant considered,
$B= 200$ MeV fm$^{-3}$ and $B= 122$ MeV fm$^{-3}$.  The maximum mass of a 
neutron
star is reduced, to 2 M$_\odot$ for a bag constant $B= 200$ MeV fm$^{-3}$ and
to 1.9 M$_\odot$ for $B= 122$ MeV fm$^{-3}$.

\begin{figure}
\centering
\epsfig{figure=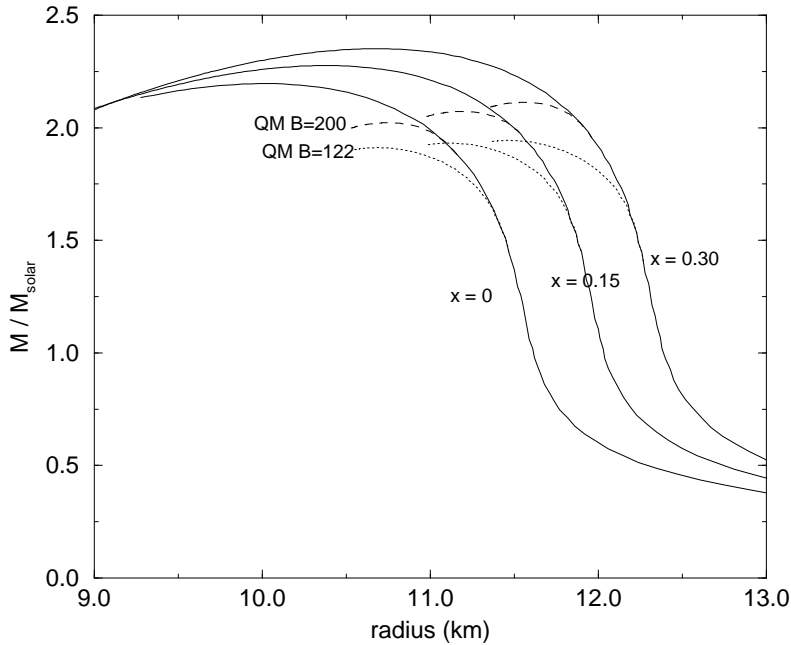, width=12cm}
\caption{Mass of neutron star plotted against radius for
force models with different three-body parts corresponding to
the values $ 0, 0.15$ and $0.30$ of the variable $x$ defined in
the text.  The dotted and dashed lines indicate the effect of matching the 
low-density equation of state 
onto a high-density one for free quarks for two values of the bag constant.}
\label{fig5}
\end{figure}

Another important question is how sensitive results are to the assumed form
of the three-body interaction.  As we have stressed earlier, the
information available empirically is scanty.  In our earlier discussion we
remarked that the three-body force was made up of two parts, a long-range
one, which we denote by $v^l_{ijk}$, and a short range one $v^s_{ijk}$.  It 
turns out that the contributions to nuclear binding energies of the long and 
short range
parts are in the same ratio for $^3$H and $^4$He,
the two light nuclei used to fit parameters, the long-range contribution being 
$-2.2$ times the short-range one.
  Thus data on these nuclei gives no
information about the relative strengths of the long and short range parts
of the interaction since any three-body interaction of the form  
\begin{equation}
v_{ijk}(x)= (1+0.45x)v^s_{ijk} +(1+x)v^l_{ijk}
\label{usx}
\end{equation}
will give an equally good fit to the properties of these two nuclei for an
arbitrary value of the parameter $x$, provided the changes can be calculated to 
linear order in $x$.  However,
the properties of symmetric nuclear matter do depend on $x$,
especially at densities higher than those of the light nuclei.
The values appropriate for the parameter $x$ are such that the
modifications that would be needed to make symmetric nuclear matter 
saturate correctly ($E = -16\; $MeV at $\rho = 0.16\; $fm$^{-3}$)
are negative ({\it ie} the energies obtained do represent an upper bound,
as is necessary for a variational calculation),
and that they are no larger than 5 MeV, a generous estimate of the
errors incurred in the variational many-body calculations.   This limits
$x$ to the range 0 to 0.3, small enough that the first-order 
treatment of the $x$ contribution is reasonably accurate.
The neutron-star analysis for the cases $x = 0$ (effectively
the same as the A18+$\delta v_b$+UIX$^*$ of \cite{apr} except
for the adjustment for saturation), and $x = 0.15$ and $0.30$,
are shown in Fig.\ref{fig5}.
The maximum mass depends on the value of $x$, and now extends from
2.20 M$_\odot$ to 2.35 M$_\odot$.   This result is merely an
estimate of the tolerance permitted by the reliance on 
fitting the  three-body interaction to the ground-state energies of only $^3$H
and $^4$He.   A new three-nucleon interaction 
that will provide much improved fits to properties of a wider 
selection of light nuclei is under development \cite{pieper}.  

\section{Concluding remarks}

There are many areas in the physics of neutron stars that we have not addressed 
in this paper.  One of 
these is the effect that possible phase transitions in the core could have on 
observable properties of 
neutron stars.  This is reviewed in a recent article by Heiselberg and 
Hjorth-Jensen \cite{hh}, who also 
investigate the effects of different interpolations between assumed low- and 
high-density equations of 
state. We have also not considered the question of what of importance to neutron 
star physics can be 
learned from laboratory experiments with heavy-ion beams.  This promises to be a 
fruitful approach, and 
the development of understanding of the physics of heavy-ion collisions that is 
a necessary prerequisite 
for constructing models of dense matter is well underway \cite{gerry}.   In 
addition we have not treated 
the advances that have occurred in the calculation of neutrino emission rates 
and transport properties.

While many mysteries remain, the advances on a number of different fronts 
(observations of neutron stars, 
nuclear experiment, and theory) are leading to an increasing understanding of 
the properties of neutron 
stars.

We are grateful to Henning Heiselberg and Fred Lamb for helpful discussions.  
This work is
supported in part by NSF grants AST 96-18524 and PHY 98-00978.

\end{document}